\documentclass[conference]{IEEEtran}

\usepackage{amsmath,amssymb}
\usepackage{graphicx}
\usepackage{booktabs}
\usepackage{tabularx}
\usepackage{url}
\usepackage[hidelinks]{hyperref}
\usepackage{balance}

\graphicspath{{figures/}}

\begin{document}

\title{SecureBank\texttrademark{}: A Financially-Aware Zero Trust Architecture for High-Assurance Banking Systems}

\author{
\IEEEauthorblockN{Paulo Fernandes Biao}
\IEEEauthorblockA{
Biaotech.dev\\
Email: \texttt{paulo@biaotech.dev}}
}

\maketitle

\begin{abstract}
Zero Trust architectures strengthen continuous verification and least-privilege enforcement, but general-purpose implementations do not explicitly treat financial exposure and transaction-level operational cost as first-class decision inputs. This paper presents SecureBank\texttrademark{}, a financially-aware Zero Trust architecture that combines identity and device trust, contextual consistency, financial risk, business-aligned segmentation, and policy actions such as allow, step-up verification, and block.

The empirical study validates the architecture's financial-risk decision component rather than claiming end-to-end production validation of the complete architecture. A public credit-card fraud dataset containing 284,807 transactions was divided into independent training, validation, and test partitions using a 60/20/20 stratified design. A class-weighted logistic regression model produced a risk score. The operational threshold was selected only on the validation partition using an asymmetric cost function in which a false negative was assigned a cost 50 times greater than a false positive, and the frozen threshold was then evaluated once on the independent test partition.

The final test set contained 56,962 transactions and 99 fraud cases. At the validation-selected risk-score threshold of 0.9991, the model achieved ROC-AUC 0.9664, average precision 0.7119, precision 71.03\%, recall 76.77\%, F1-score 0.7379, and a false-positive rate of 0.0545\%. These results provide reproducible component-level evidence that financially asymmetric operating costs can be integrated into a Zero Trust decision process while keeping the scope of the empirical claims explicit.
\end{abstract}


\section{Introduction}

Financial institutions process high-volume transactions across mobile applications, APIs, cloud services, employee systems, and third-party integrations. In this environment, a decision that is technically valid from an identity or network perspective may still create unacceptable financial exposure. Zero Trust Architecture (ZTA) addresses implicit trust by emphasizing continuous verification, resource-oriented policy enforcement, and least privilege~\cite{NIST800207}. However, its implementation guidance is intentionally general and does not define a banking-specific method for combining access context with transaction risk and asymmetric financial consequences.

SecureBank\texttrademark{} addresses this gap by treating financial risk as a direct input to the policy decision process. The architecture combines identity trust, device posture, contextual consistency, transaction-level risk, service criticality, and explicit response actions. Its purpose is not to replace fraud detection, identity management, segmentation, or security orchestration platforms, but to coordinate their signals through a financially-aware decision layer.

This paper makes four contributions:

\begin{enumerate}
\item It defines a financially-aware Zero Trust architecture for banking and payment environments.
\item It formulates a composite trust-risk decision model that maps contextual and financial signals to explainable policy actions.
\item It introduces a cost-sensitive threshold-selection procedure that separates model training, threshold calibration, and final testing.
\item It reports reproducible empirical evidence for the financial-risk decision component on a highly imbalanced public fraud dataset.
\end{enumerate}

The empirical claims are deliberately limited. The public dataset contains anonymized transaction features but does not contain real identity, device, network-segmentation, or SOAR telemetry. Consequently, the experiment validates the risk-scoring and cost-sensitive decision component, while the wider architecture remains a design framework requiring deployment-oriented evaluation.

\section{Background and Research Gap}

\subsection{Zero Trust and Service-Centric Enforcement}

NIST SP 800-207 defines ZTA as a shift from static perimeter defenses toward policies centered on users, assets, and resources~\cite{NIST800207}. NIST SP 800-207A further describes granular application-level access control in cloud-native environments~\cite{NIST800207A}. Google's BeyondProd similarly emphasizes service identity, workload protection, and security controls that follow cloud-native services rather than a trusted network perimeter~\cite{BeyondProd}.

These frameworks provide strong architectural foundations but are not designed as transaction-cost models. In banking, policy decisions may need to consider expected financial loss, fraud-investigation workload, settlement exposure, regulatory impact, and the cost of interrupting legitimate activity.

\subsection{Financial Resilience and Operational Governance}

Financial institutions operate under security and resilience requirements that include payment-data protection, technology-risk governance, incident handling, and digital operational resilience. PCI DSS v4.0.1 defines baseline requirements for protecting payment account data~\cite{PCIDSS}. The FFIEC Architecture, Infrastructure, and Operations booklet addresses technology architecture and operational risk management in financial institutions~\cite{FFIECAIO}. The European Union's Digital Operational Resilience Act establishes requirements for ICT risk and operational resilience in the financial sector~\cite{DORA}. These sources motivate an architecture in which technical controls are connected to operational impact rather than evaluated in isolation.

\subsection{Fraud Detection Under Class Imbalance}

Credit-card fraud detection is characterized by severe class imbalance, non-stationarity, delayed verification, and asymmetric error costs~\cite{DalPozzolo2014,DalPozzolo2018}. In such settings, ROC-AUC alone can obscure operational performance, and precision-recall analysis is particularly informative~\cite{Saito2015}. Cost-sensitive learning provides a principled way to represent the fact that different errors impose different consequences~\cite{Elkan2001}.

The research gap addressed here lies at the intersection of these areas: ZTA supplies continuous decision and enforcement principles, while fraud detection supplies transaction-level risk evidence and asymmetric operating costs. SecureBank integrates them as a single policy-decision workflow.

\section{SecureBank Architecture}

\subsection{Architectural Components}

SecureBank is organized around five logical components:

\begin{enumerate}
\item \textbf{Signal ingestion}: identity, device, session, transaction, service-criticality, and threat-intelligence signals.
\item \textbf{Financial-risk engine}: estimation of transaction risk and potential operational impact.
\item \textbf{Trust-risk policy decision point}: combination of trust and risk evidence into an explainable decision.
\item \textbf{Policy enforcement}: allow, step-up verification, rate limit, isolate, block, or escalate.
\item \textbf{Feedback and governance}: logging, analyst review, model monitoring, audit evidence, and policy recalibration.
\end{enumerate}

The architecture is compatible with a resource-oriented ZTA: the policy decision point evaluates each request using current evidence, while enforcement occurs close to the protected application, API, or workflow.

\subsection{Composite Trust-Risk Model}

Let the composite decision score at time $t$ be

\begin{equation}
\Theta_t = w_i I_t + w_d D_t + w_c C_t - w_r R_t,
\label{eq:theta}
\end{equation}

where $I_t$ is identity trust, $D_t$ is device posture, $C_t$ is contextual consistency, and $R_t$ is financial risk. The weights are non-negative and should be institution-specific.

An anomaly may reduce accumulated trust according to

\begin{equation}
\Theta_{t+1} = \Theta_t e^{-\lambda A_t},
\label{eq:decay}
\end{equation}

where $A_t$ represents anomaly intensity and $\lambda$ controls the decay rate. These equations express the architectural logic; they are not claimed to have been fully validated by the public fraud dataset.

\subsection{Financial Impact and Policy Actions}

For a transaction or access request $x$, the architecture can represent an operational risk function as

\begin{equation}
R(x) = g\!\left(s(x),\,a(x),\,c(x),\,e(x)\right),
\end{equation}

where $s(x)$ is a predictive risk score, $a(x)$ is transaction amount or exposure, $c(x)$ is service criticality, and $e(x)$ represents contextual or regulatory impact. Policy bands then map the combined evidence to actions:

\begin{equation}
\pi(x)=
\begin{cases}
\text{allow}, & R(x)<\tau_1,\\
\text{step-up}, & \tau_1 \leq R(x)<\tau_2,\\
\text{block/escalate}, & R(x)\geq\tau_2.
\end{cases}
\end{equation}

In the empirical study, only the binary fraud-risk decision and its cost-sensitive threshold are evaluated. Multi-action enforcement remains an architectural extension.

\section{Threat Model and Scope}

The architecture considers credential compromise, session hijacking, device anomalies, API abuse, insider misuse, and transaction fraud. Adversaries may possess valid credentials or partial knowledge of business workflows. SecureBank assumes that upstream systems can provide trustworthy telemetry and that enforcement points can execute policy decisions.

The empirical dataset does not contain direct observations of credentials, device posture, network movement, analyst actions, or security orchestration. Therefore, it cannot validate identity adaptation, contextual micro-segmentation, or automated incident response. Those capabilities are architectural claims, not empirical findings of the present experiment.

\section{Empirical Methodology}

\subsection{Dataset}

The empirical evaluation uses the public European credit-card fraud dataset distributed by the Machine Learning Group of Universit\'e Libre de Bruxelles~\cite{ULBDataset}. It contains 284,807 transactions, including 492 fraud cases, for a fraud prevalence of approximately 0.173\%. Features $V1$--$V28$ are anonymized numerical components, while \texttt{Time} and \texttt{Amount} are provided separately.

Because the semantic meaning of $V1$--$V28$ is not disclosed, the experiment does not relabel them as identity, device, or context attributes. The features are used only as predictive transaction variables.

\subsection{Data Partitioning}

A stratified two-stage split creates independent partitions:

\begin{itemize}
\item 60\% for model training;
\item 20\% for threshold selection;
\item 20\% for final testing.
\end{itemize}

The resulting counts are shown in Table~\ref{tab:splits}. The final test partition is not used during model fitting or threshold calibration.

\begin{table}[t]
\centering
\caption{Stratified dataset partitions.}
\label{tab:splits}
\begin{tabular}{lrr}
\toprule
Partition & Transactions & Fraud cases\\
\midrule
Training & 170,884 & 295\\
Validation & 56,961 & 98\\
Test & 56,962 & 99\\
\bottomrule
\end{tabular}
\end{table}

\subsection{Model and Preprocessing}

A logistic regression classifier with balanced class weights is used as an interpretable empirical baseline. Missing-value imputation and feature standardization are contained in a single training pipeline and are fitted only on the training partition. The model then produces continuous scores for the validation and test partitions.

Because balanced class weighting alters the relationship between model output and observed prevalence, the output is described as a \emph{risk score}, not as a calibrated fraud probability.

\subsection{Cost-Sensitive Threshold Selection}

Let $FP(\tau)$ and $FN(\tau)$ denote false positives and false negatives at threshold $\tau$. The validation objective is

\begin{equation}
C(\tau)=c_{FP}FP(\tau)+c_{FN}FN(\tau),
\label{eq:cost}
\end{equation}

with $c_{FP}=1$ and $c_{FN}=50$. These values are an explicit experimental assumption representing the greater cost of undetected fraud. They are not universal banking cost estimates.

Every distinct validation score is considered as a candidate threshold. The threshold minimizing $C(\tau)$ is frozen and applied once to the independent test partition.

\subsection{Evaluation Metrics}

The evaluation reports ROC-AUC, average precision (PR-AUC), precision, recall, F1-score, false-positive rate, confusion-matrix counts, and weighted operational cost. PR-AUC is included because the positive class is highly imbalanced~\cite{Saito2015}.

\section{Empirical Results}

\subsection{Validation-Based Threshold Selection}

The minimum validation cost was obtained at a risk-score threshold of

\begin{equation}
\tau^\ast = 0.999133655312.
\end{equation}

At this threshold, the validation partition produced 85 true positives, 34 false positives, and 13 false negatives, corresponding to a weighted validation cost of 684.

\begin{figure}[t]
\centering
\includegraphics[width=\linewidth]{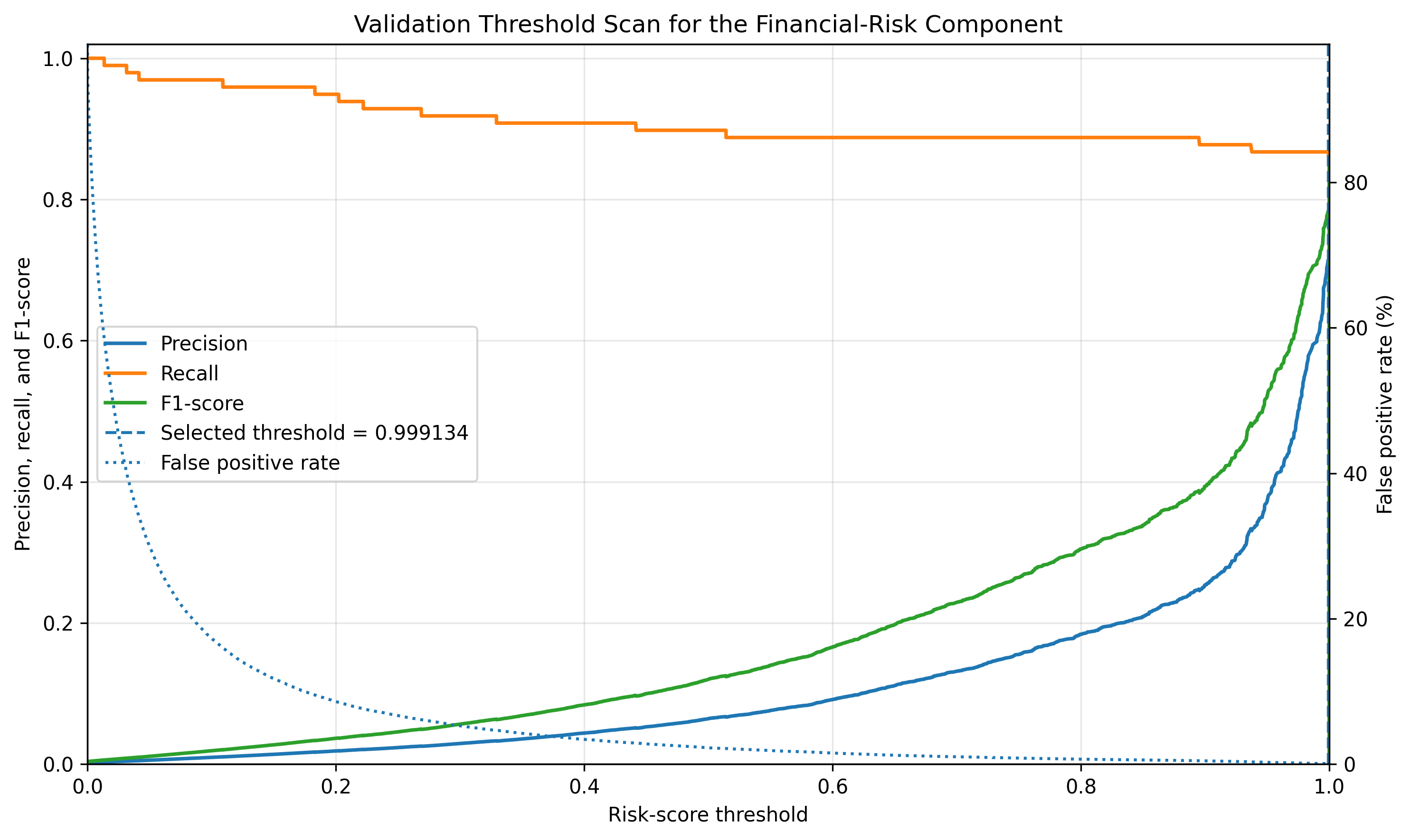}
\caption{Validation-only threshold scan. The vertical marker indicates the threshold selected before the independent test evaluation.}
\label{fig:threshold-metrics}
\end{figure}

\begin{figure}[t]
\centering
\includegraphics[width=\linewidth]{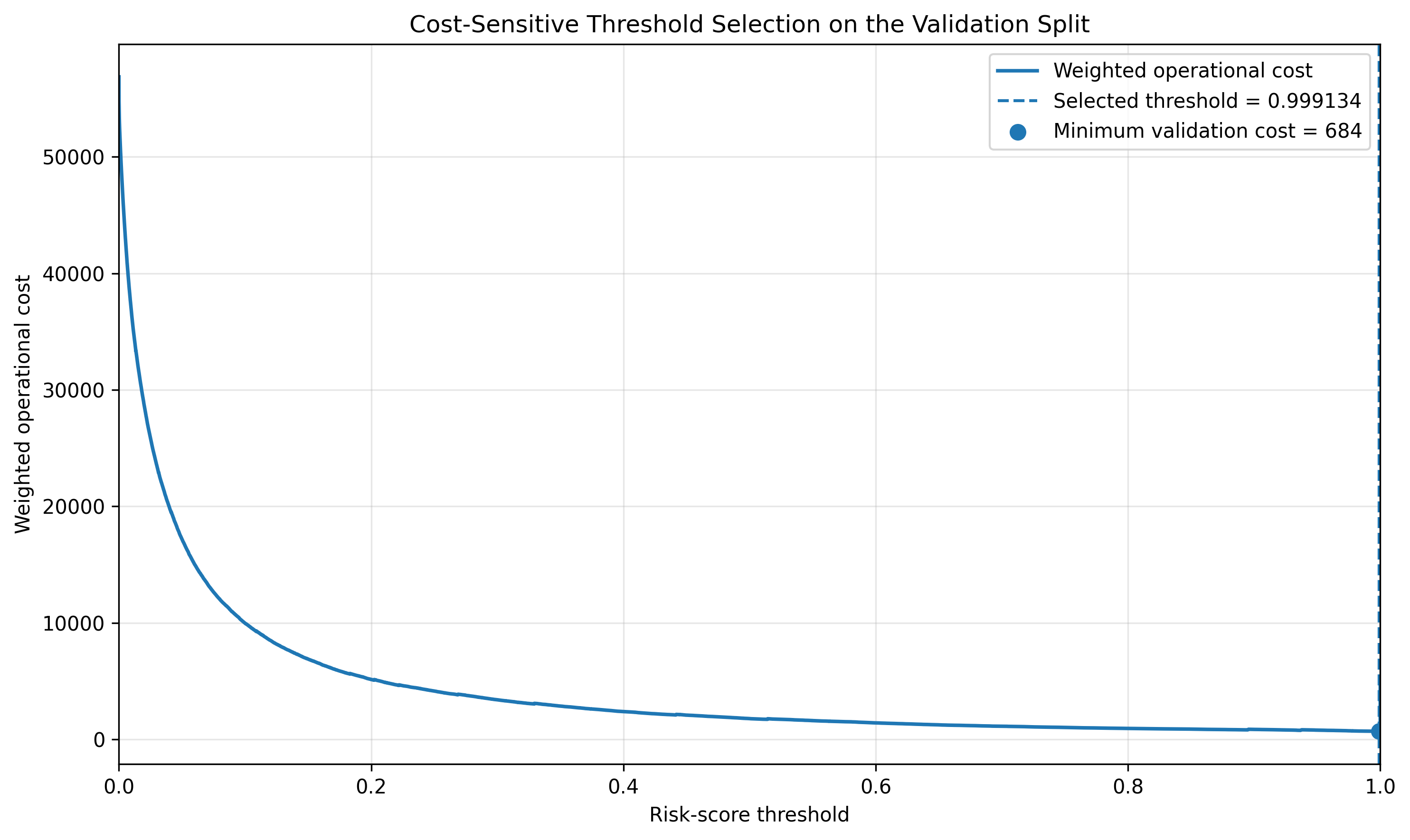}
\caption{Weighted operational cost on the validation partition using $c_{FP}=1$ and $c_{FN}=50$.}
\label{fig:threshold-cost}
\end{figure}

Figure~\ref{fig:threshold-metrics} shows the validation trade-off among precision, recall, F1-score, and false-positive rate. Figure~\ref{fig:threshold-cost} shows the cost objective used to choose the frozen threshold.

\subsection{Independent Test Evaluation}

The frozen threshold was applied to 56,962 previously unused test transactions. Table~\ref{tab:results} reports the final results.

\begin{table}[t]
\centering
\caption{Independent test results at the frozen threshold.}
\label{tab:results}
\begin{tabular}{lr}
\toprule
Metric & Value\\
\midrule
Risk-score threshold & 0.999133655312\\
Test transactions & 56,962\\
Fraud cases & 99\\
True positives & 76\\
False positives & 31\\
True negatives & 56,832\\
False negatives & 23\\
Precision & 71.03\%\\
Recall & 76.77\%\\
F1-score & 0.7379\\
False-positive rate & 0.0545\%\\
ROC-AUC & 0.9664\\
Average precision & 0.7119\\
Weighted test cost & 1,181\\
\bottomrule
\end{tabular}
\end{table}

\begin{figure}[t]
\centering
\includegraphics[width=\linewidth]{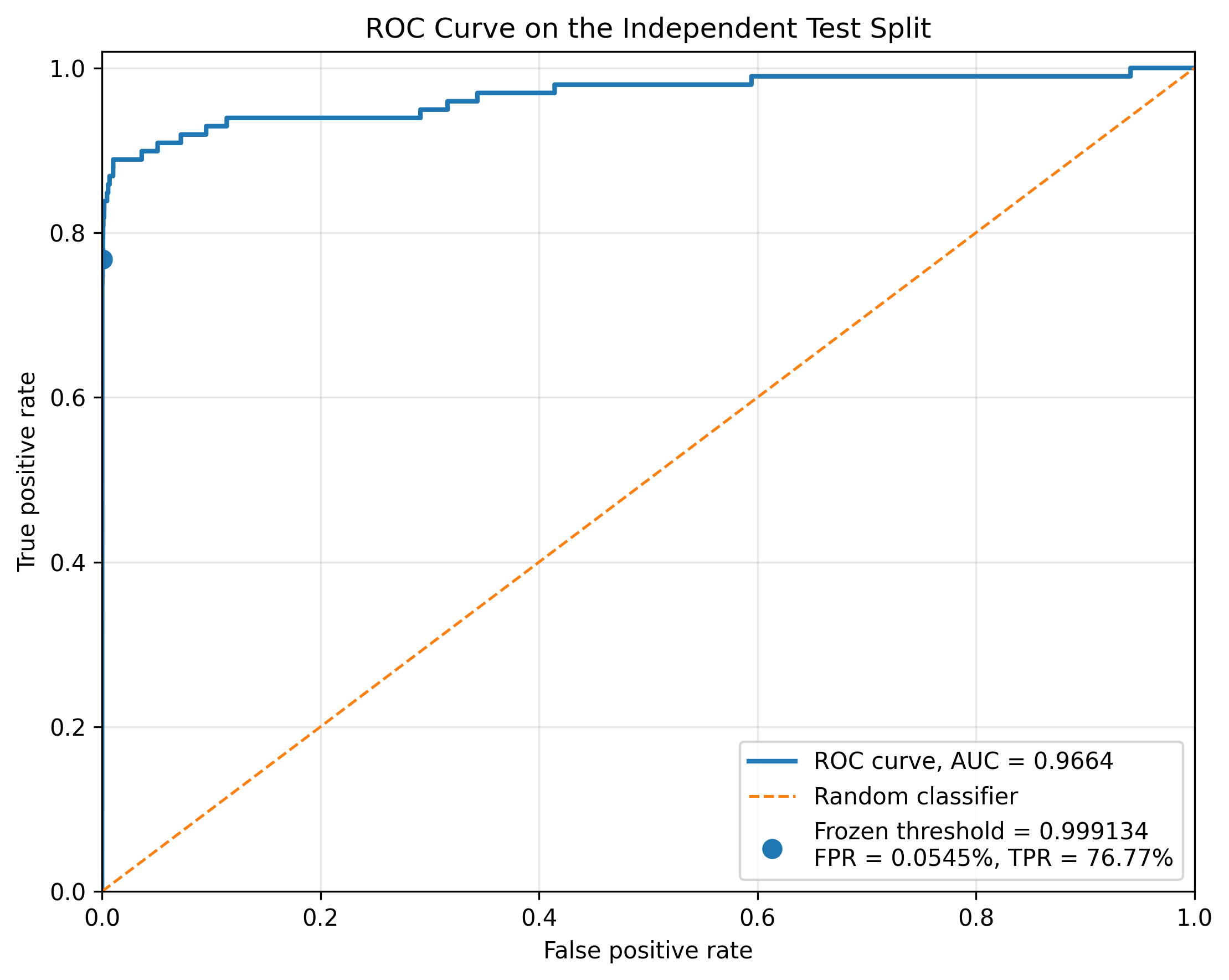}
\caption{ROC curve on the independent test partition. The selected operating point was fixed using the validation partition.}
\label{fig:roc}
\end{figure}

\begin{figure}[t]
\centering
\includegraphics[width=\linewidth]{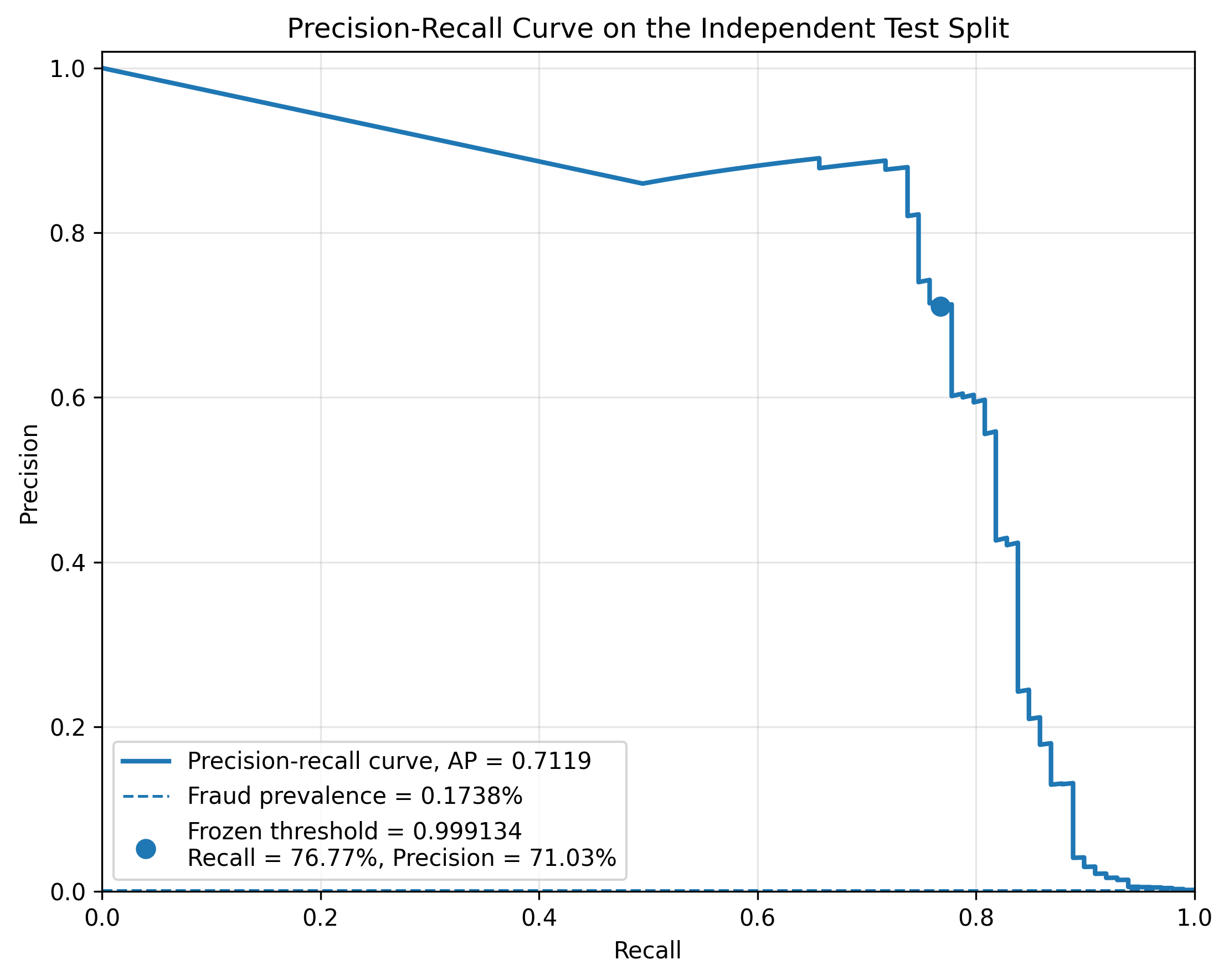}
\caption{Precision-recall curve on the independent test partition, including the frozen operating point.}
\label{fig:pr}
\end{figure}

Figure~\ref{fig:roc} presents the ROC curve on the independent test partition. Figure~\ref{fig:pr} presents the corresponding precision-recall curve and the frozen operating point.

The final precision of 71.03\% means that 76 of the 107 flagged test transactions were fraudulent. The recall of 76.77\% means that 76 of the 99 fraud cases were detected. The false-positive rate was 0.0545\%, corresponding to 31 legitimate transactions among 56,863 legitimate test cases.

\section{Discussion}

The results support three conclusions.

First, the risk model has strong ranking ability on the held-out data, as indicated by ROC-AUC 0.9664 and average precision 0.7119. Average precision is particularly relevant because the fraud prevalence is only 0.1738\% in the test partition.

Second, separating threshold calibration from final testing materially strengthens the validity of the operating-point results. The threshold is not selected after observing the test labels.

Third, the experiment demonstrates a financially-aware decision principle: the operating point can be selected according to explicit error costs rather than a default score threshold. The selected cutoff is specific to the model, data split, and 50:1 cost assumption. It should not be interpreted as a universal threshold or as a calibrated 99.91\% fraud probability.

The empirical results do not demonstrate that the entire SecureBank architecture has been deployed or validated in a production bank. They validate one component: transforming transaction evidence into a cost-sensitive risk decision that can be consumed by a Zero Trust policy engine.

\section{Limitations and Threats to Validity}

The study has several limitations.

\begin{itemize}
\item The dataset covers only two days of transactions and may not represent long-term concept drift.
\item The anonymized PCA components prevent semantic interpretation as identity, device, or session features.
\item The logistic regression model is a baseline rather than a state-of-the-art model comparison.
\item The 50:1 cost ratio is an experimental assumption and should be replaced by institution-specific estimates in deployment studies.
\item Balanced class weights produce useful ranking scores but not necessarily calibrated probabilities.
\item The wider architecture, including multi-action enforcement, micro-segmentation, SOAR integration, latency, and regulatory workflows, is not validated by this dataset.
\item Repeated evaluation across alternative random splits or temporal partitions is needed to quantify uncertainty around the reported point estimates.
\end{itemize}

\section{Reproducibility}

The implementation uses fixed random seeds, a documented 60/20/20 stratified split, a preprocessing-and-model pipeline fitted only on training data, exhaustive threshold evaluation on validation scores, and one-time evaluation on the test partition. The repository should include the training script, threshold-selection script, plotting scripts, dependency file, and machine-readable result files. The raw transaction dataset should be obtained from its original distribution page rather than committed to the repository.

\section{Future Work}

Future work will extend the evaluation in four directions:

\begin{enumerate}
\item compare logistic regression with tree ensembles and calibrated models under identical splits;
\item repeat the experiment across multiple seeds and temporal partitions with confidence intervals;
\item evaluate model calibration and institution-specific cost functions;
\item integrate identity, device, API, segmentation, and incident-response telemetry in a deployment-oriented testbed.
\end{enumerate}

A separate simulation study may evaluate adaptive trust and policy-enforcement behavior, but its metrics must be defined so that directionality and interpretation are unambiguous before quantitative claims are added to the manuscript.

\section{Conclusion}

SecureBank proposes a financially-aware extension of Zero Trust decision-making for banking and payment systems. The architecture combines trust evidence with transaction risk and operational impact, while the empirical study isolates and validates its financial-risk decision component.

Using a reproducible training-validation-test design, the component achieved ROC-AUC 0.9664, average precision 0.7119, precision 71.03\%, recall 76.77\%, F1-score 0.7379, and a false-positive rate of 0.0545\% on an independent test partition. The results show that asymmetric financial costs can guide a defensible operating threshold without using the final test set for calibration.

The contribution is therefore architectural and component-level: SecureBank provides a framework for connecting financially-aware risk scoring to Zero Trust policy enforcement, while full system effectiveness remains a subject for broader deployment and longitudinal evaluation.

\end{document}